\documentclass[letterpaper]{article} 
\usepackage{aaai2026}  
\usepackage{times}  
\usepackage{helvet}  
\usepackage{courier}  
\usepackage[hyphens]{url}  
\usepackage{graphicx} 
\usepackage{amsfonts} 
\usepackage{natbib}  
\usepackage{caption} 
\usepackage{pifont}       
\usepackage{algorithm}
\usepackage{algorithmic}
\usepackage{multirow}
\usepackage{amsmath}
\usepackage{booktabs}
\usepackage[table]{xcolor}
\usepackage{makecell}
\newcommand{\cmark}{\ding{51}}
\newcommand{\xmark}{\ding{55}}

\definecolor{cvprblue}{rgb}{0.21,0.49,0.74}
\usepackage{newfloat}
\usepackage{listings}
\DeclareCaptionStyle{ruled}{labelfont=normalfont,labelsep=colon,strut=off} 
\floatstyle{ruled}
\newfloat{listing}{tb}{lst}{}
\floatname{listing}{Listing}
\title{\texttt{Align$^3$GR}: Unified Multi-Level Alignment for LLM-based Generative Recommendation}
\author{
    Wencai Ye\equalcontrib,
    Mingjie Sun\equalcontrib,
    Shuhang Chen\equalcontrib,
    Wenjin Wu\thanks{Corresponding Author.},
    Peng Jiang
}
\affiliations{
    Kuaishou Technology, China\\
    {\{yewencai, sunmingjie, chenshuhang, wuwenjin, jiangpeng\}}@kuaishou.com
}

\usepackage{bibentry}

\begin{document}
\maketitle

\begin{abstract}
Large Language Models (LLMs) demonstrate significant advantages in leveraging structured world knowledge and multi-step reasoning capabilities. However, fundamental challenges arise when transforming LLMs into real-world recommender systems due to semantic and behavioral misalignment. 
To bridge this gap, we propose \textbf{Align$^3$GR}, a novel framework that unifies token-level, behavior modeling-level, and preference-level alignment. Our approach introduces: Dual tokenization fusing user-item semantic and collaborative signals. Enhanced behavior modeling with bidirectional semantic alignment. Progressive DPO strategy combining self-play (SP-DPO) and real-world feedback (RF-DPO) for dynamic preference adaptation. Experiments show Align$^3$GR outperforms the SOTA baseline by +17.8\% in Recall@10 and +20.2\% in NDCG@10 on the public dataset, with significant gains in online A/B tests and full-scale deployment on an industrial large-scale recommendation platform.
\end{abstract}

\section{Introduction}
Recommender systems (RS) are essential infrastructures in modern digital platforms such as e-commerce~\cite{e-com}, video streaming~\cite{vedio}, and social media~\cite{social-media}. With the rapid advancement of large language models (LLMs), researchers have explored two primary paradigms for integrating LLMs into recommender systems. Initially, LLMs have been employed to enhance traditional discriminative recommenders, for example, by providing improved content and user understanding~\cite{wu2023survey}, or enabling query rewriting and reasoning~\cite{ye2023informative}. More recently, LLMs have been developed as standalone generative recommenders, where the model directly outputs recommended items in an end-to-end manner~\cite{tiger, lc-rec}.
This shift from augmentation to full replacement brings forth a fundamental challenge: \textit{How can LLMs be truly transformed into recommender systems?}
\begin{figure}[h]
    \centering
    \includegraphics[width=1.0\linewidth]{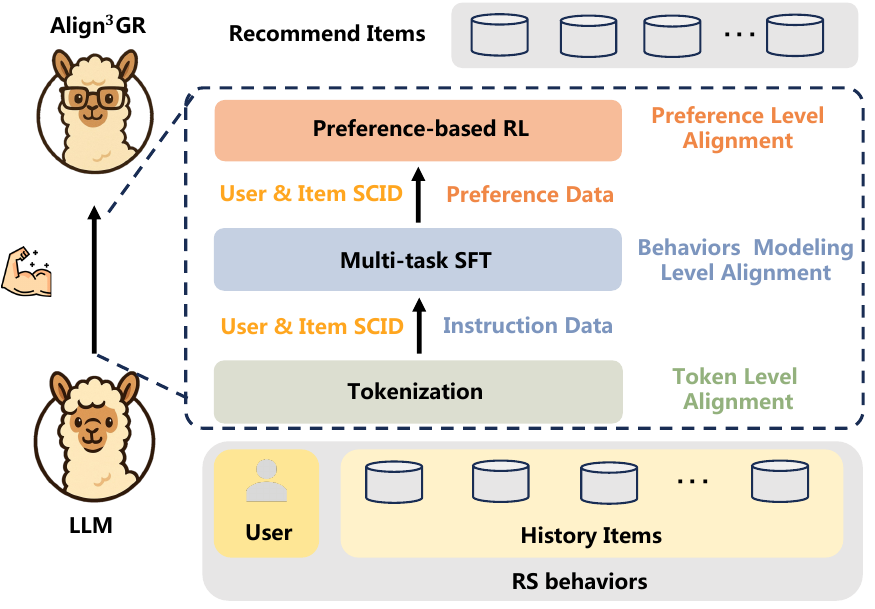}
    \caption{LLM-to-Recommendation Alignment Pipeline.}
    \label{fig:fig1}
\end{figure}
The key is effectively aligning the fundamental gap between the pre-trained LLMs and the personalized recommendation system~\cite{seLLaRec}: On the one hand, language modeling is primarily concerned with the semantic information with next-token prediction (NTP); on the other hand, recommender systems tend to model users' implicit preferences based on their interaction behavior information~\cite{goal}. 
Recent research efforts~\cite{li2025semantic, eager} have attempted to bridge this gap. We systematically summarized them into a unified alignment pipeline spanning three indispensable levels: tokenization, multi-task supervised fine-tuning (SFT), and preference-based reinforcement learning (RL).
 \begin{figure*}[t]
    \centering
    \includegraphics[width=1\linewidth]{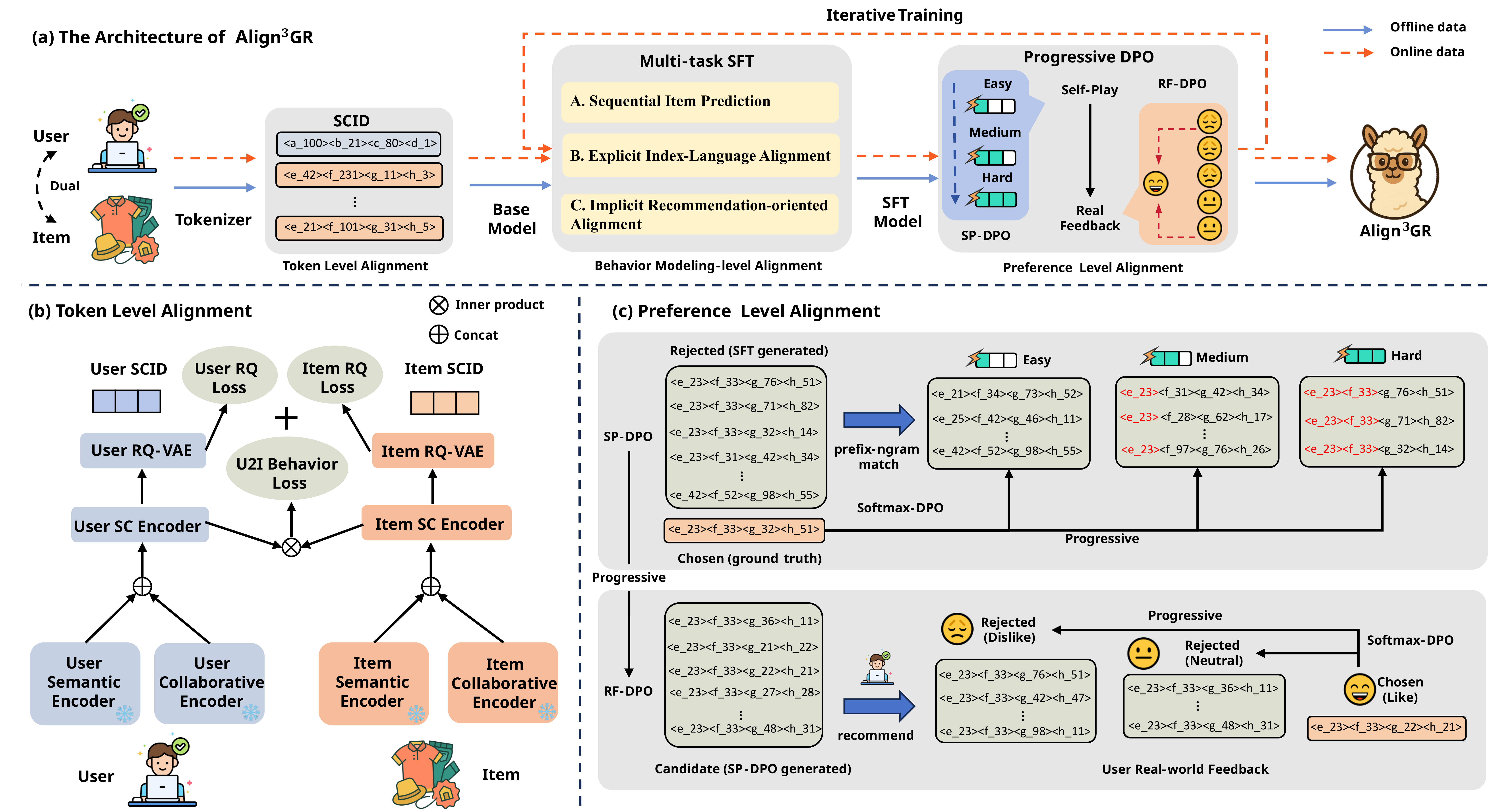}
    \caption{(a) The architecture of \textbf{Align$^3$GR}, a unified multi-level alignment framework for generative recommendation, which integrates hierarchical dual SCID, multi-task SFT, and progressive DPO. (b) Token-level alignment is achieved through user-item dual SC encoders and RQ-VAEs. (c) Preference-level alignment is accomplished via progressive SP-DPO and RF-DPO.}
    \label{fig:fig2}
\end{figure*}
First, tokenization serves as the foundation by transforming user and item information into compact and expressive token representations~\cite{sidsurvey}, and EAGER~\cite{eager} and DAS~\cite{das} have incorporated collaborative signals at this stage to better bridge the gap between language modeling and downstream NTP tasks that require user-item interaction modeling. Second, the objective of the SFT stage is to enable general-purpose LLMs to comprehend the data structures and user behavior patterns inherent to recommendation tasks, thereby equipping the model with initial recommendation capabilities~\cite{eager-llm}. Third, preference-based RL is essential for aligning model outputs with real user interests and business objectives, thereby enhancing the personalization of recommendations~\cite{onerec}.
Despite their merits, previous methods often treat user and item information independently during tokenization, neglecting the collaborative and semantic dependencies between them. Such isolated modeling fails to capture the mutual influences critical for comprehensive preference learning and has been shown to result in suboptimal recommendation performance~\cite{collm}. On the other hand, aligning LLMs with real-world user preferences and business objectives presents additional challenges~\cite{rf1,rf2}. Although recent preference-based RL methods like DPO~\cite{dpo1,cdpo} attempt to incorporate user preference signals into the RS, they typically rely on the collection of offline data without any progressive learning mechanisms. In real-world recommendation scenarios, user behavioral preferences are inherently complex and ambiguous compared to explicit labels, increasing the difficulty for models to learn directly.
To address these challenges, we propose Align$^3$GR, a unified multi-level alignment framework that systematically integrates token-level, behavior modeling-level, and preference-level alignment. Specifically, our approach first introduces a hierarchical tokenization scheme, enabling compact and jointly optimized Semantic-Collaborative ID (SCID) for both users and items, which lays a solid foundation for subsequent SFT and preference-based RL, ensuring effective alignment throughout the entire recommendation pipeline.
Next, during behavior modeling-level alignment, we further introduce two additional tasks. We first incorporate the user's SCID into the main sequence generation task, enriching the input representation and enabling the model to leverage more comprehensive user features. Then, we introduce a bidirectional alignment task between the user’s SCID and their semantic information, which explicitly grounds the SCID tokens in their real-world semantic meanings.
Furthermore, inspired by curriculum learning~\cite{llara}, we introduce a progressive DPO strategy (easy to hard) that enables the model to improve continually by learning from preference pairs of increasing difficulty, leading to smoother convergence and more stable training. 
Building on this, we adopt the self-play DPO (SP-DPO), which involves interacting with itself to generate diverse training data, thereby performing cost-efficient adversarial bias correction on SFT models~\cite{spdpo} and alleviating the exploration bottleneck of traditional RL. However, the lack of real-world feedback during self-play limits the model’s ability to generalize to real-world scenarios. Hence, we incorporate real-world feedback DPO (RF-DPO), progressively aligning the model with actual user interests and business objectives. Specifically, the RF-DPO also follows a progressive strategy. Finally, we conduct comprehensive experiments on both public and industrial datasets to validate the effectiveness of our approach. On the Instruments dataset, Align$^3$GR outperforms the SOTA baseline by 17.8\% in Recall@10 and 20.2\% in NDCG@10, demonstrating consistent and substantial improvements over strong generative baselines. Moreover, the gains are further corroborated by real-world business metrics in industrial settings. The core contributions of this work are as follows:
\textbf{1)} We propose Align$^3$GR, a unified alignment framework that jointly optimizes token-level, behavior modeling-level, and preference-level alignment to bridge the gap between LLMs and RS.
\textbf{2)} We design a dual tokenization scheme that enables semantic and collaborative fusion at the input level. In addition, we develop a progressive DPO strategy, including SP-DPO and RF-DPO, adapted for dynamic scenario preference alignment in RS.
\textbf{3)} Comprehensive experiments on benchmark and industrial datasets demonstrate consistent performance gains across all settings, with Align$^3$GR yielding significant improvements in offline metrics and online A/B testing spanning diverse recommendation scenarios.
\section{Related Works}
\subsection{Generative Recommendation}
Generative recommendation~\cite{gr1} reframes classical retrieval as a sequence generation task, enabling models to benefit from complete semantic context and dynamically adaptable output formats. Generative approaches are different from standard embedding-based design (e.g., two-tower models with ANN search~\cite{twt1}) in that they autoregressively generate item tokens (e.g., IDs, titles, semantic IDs), enabling dynamic, context-aware recommendations. The three key technical directions are as follows: First, sequence generation retrieval (e.g., DSI~\cite{dsi1,dsi2} and GENRE~\cite{genre}) transforms retrieval into autoregressive user context token generation. Second, breakthroughs in indexing or tokenization, such as RQ-VAE~\cite{rqvae}, hierarchical k-means~\cite{kmeans}, and PQ~\cite{pq}, transform content embeddings into short discrete tokens, powering generative recommendation models like TIGER~\cite{tiger}, LC-Rec~\cite{lc-rec}, LETTER~\cite{letter}, and EAGER-LLM~\cite{eager-llm}.
\subsection{Preference Alignment of LLMs}
Alignment of LLMs with human preferences has been extensively studied in NLP, primarily through Reinforcement Learning from Human Feedback (RLHF)~\cite{rlhf2} and Direct Preference Optimization (DPO)~\cite{dpo}. While RLHF guides policy updates via reward models built from human annotations, it suffers from instability and high computational costs. DPO addresses these limitations by directly optimizing model parameters on preference data, inspiring variants such as IPO~\cite{ipo}, cDPO~\cite{cdpo}, rDPO~\cite{rdpo}, and Softmax-DPO~\cite{softmax-dpo} to tackle noise robustness and unbiased learning. However, applying these techniques to recommender systems remains challenging due to sparse, implicit preference data, which limits static offline optimization and adaptation to dynamic user interests and business objectives~\cite{onerec}. Recent advances, including curriculum learning~\cite{llara} and self-play~\cite{selfplay, spdpo}, have improved preference alignment. Notably, progressive strategies enhance robustness by organizing training from easy to hard preference pairs. 
Our work builds on these progressive strategies to bridge static preference optimization and the dynamic demands of large-scale recommender systems.
\section{Methodology}
\label{method}
\subsection{Overview of the Proposed Framework}
In this work, we propose Align$^3$GR—a unified multi-level alignment framework that systematically bridges the gap between LLMs and recommender systems. As illustrated in Figure~\ref{fig:fig2}, our framework consists of three tightly aligned stages:
First, at the token level, we introduce a novel tokenization scheme based on a user-item dual learning strategy, enabling dual-track fusion of user/item-specific semantic-collaborative features to generate hierarchical discrete SCID, while maintaining a compact token space and minimizing computational overhead during both training and inference.
Second, at the behavior modeling level, we design an enhanced multi-task SFT based on LC-Rec~\cite{lc-rec} by not only incorporating the SCID of users but also explicitly aligning these SCIDs with their semantic information through the user alignment task during explicit index-language alignment. 
Finally, at the preference alignment level, we draw inspiration from curriculum learning to propose a progressive DPO strategy (easy to hard), combining self-play DPO (SP-DPO) for continual self-improvement with real-world feedback DPO (RF-DPO), while simultaneously addressing preference learning from both generative performance and real user feedback perspectives. Together, these three levels form a coherent alignment pipeline that enables our LLM-based GR model to achieve both high-quality personalization and robust adaptation in large-scale dynamic recommendation.
\subsection{Token-level Alignment: Dual SCID Tokenization}
Current tokenization methods for generative recommendation primarily encode items while overlooking user structure modeling. Although some incorporate user representations, these are seldom co-optimized with item embeddings, resulting in suboptimal user-item alignment and representations that lack critical collaborative signals for recommendation.
We argue that a truly effective tokenization scheme should jointly encode both users and items, leveraging their respective semantic and collaborative signals, and must leverage a unified, co-optimized framework to learn mutually aligned, expressive representations.
\begin{figure}[t]
    \centering
    \includegraphics[width=1.0\linewidth]{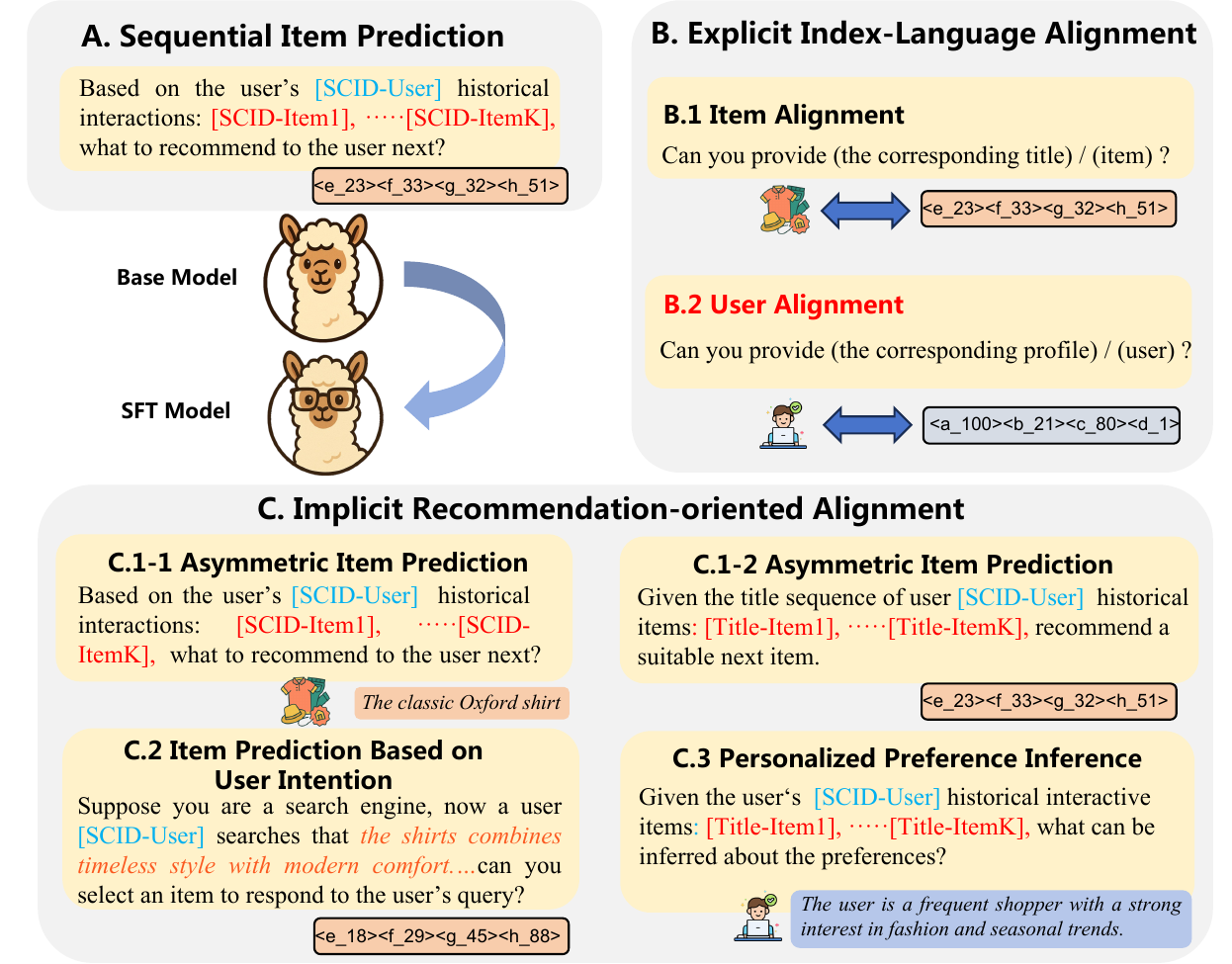}
    \caption{Behavior Modeling-level Alignment.}
    \label{fig:fig3}
\end{figure}
Dual SCID Tokenization addresses these gaps by integrating semantic and collaborative features for users and items within a unified dual learning framework (Figure~\ref{fig:fig2}b). Specifically, we first extract both semantic features (e.g., profiles or descriptions) and collaborative features (e.g., behavioral patterns) for users and items. These features are processed by dedicated encoders: a frozen semantic encoder (initialized with LMs like T5~\cite{t5}) captures textual representations, while a frozen collaborative encoder (e.g., DIN~\cite{din}) models behavioral dynamics. The resulting semantic and collaborative embeddings are then concatenated and passed through a hybrid Semantic-Collaborative (SC) Encoder (e.g., MLP), which integrates both information types to produce unified SC embeddings $(\mathbf{SC}{\text{u}},\ \mathbf{SC}{\text{i}})$. Finally, we quantize these unified embeddings into SCID using RQ-VAE~\cite{rqvae}.
The training objective consists of two main components. First, at the embedding level, we optimize a sampled-softmax user-item behavior loss to enhance alignment between user and item SC embeddings.
$\mathcal{L}_{\text{U2I}}$, defined as follows:
\begin{small}
  \begin{align}
\mathcal{L}_{\text{U2I}} = -\frac{1}{|\mathcal{B}|} \sum_{(u,i^+) \in \mathcal{B}} \left[ \log \frac{\exp(\mathbf{u}_u^\top \mathbf{v}_{i^+})}{\exp(\mathbf{u}_u^\top \mathbf{v}_{i^+}) + \sum_{j \in \mathcal{N}_u} \exp(\mathbf{u}_u^\top \mathbf{v}_j)} \right]
\end{align}  
\end{small}
where $\mathcal{B}$ is the batch size, and $\mathbf{u}$ and $\mathbf{v}$ represent $\mathbf{SC}{\text{u}}$ and $\mathbf{SC}{\text{i}}$, the positive examples are user-item interaction behaviors. $\mathcal{N}_u$ is the set of negative samples, which are randomly sampled within the batch.
Second, the overall joint loss combines the U2I behavior loss and quantization losses from user-specific and item-specific RQ-VAE as:
\begin{small}
\begin{equation}
\mathcal{L} = \alpha \cdot \mathcal{L}_{\text{U2I}} + \gamma \cdot \left( \mathcal{L}_{\text{User RQ}} + \mathcal{L}_{\text{Item RQ}} \right)
\end{equation}
\end{small}
where $\mathcal{L}_{\text{User RQ}}$ and $\mathcal{L}_{\text{Item RQ}}$ denote the reconstruction and quantization losses for user and item embeddings, and $\alpha, \gamma$ are trade-off hyperparameters.
In practice,  we first set $\alpha=1, \gamma=0$ and optimize $\mathcal{L}_{\text{U2I}}$ to stabilize behavior alignment and ensure thorough learning of the SC Encoder (monitored via AUC). Once stabilized, we switch to $\alpha=0.1, \gamma=1$ to focus on optimizing the quantization losses. During inference, user and item modules are deployed separately, each generating their own SCID for downstream use.
This design not only compresses the token space but also enables more effective downstream multi-task SFT and preference-based RL by preserving collaborative relationships throughout the model pipeline.
\subsection{Behavior Modeling-level Alignment: Multi-task SFT}
After obtaining quantized SCIDs for each user and item through alignment tokenization, we continue to enhance the generative and semantic alignment capabilities of the LLM in the new token space. Following previous work~\cite{lc-rec}, we first expand the LLM vocabulary to include the SCID tokens of the user and the item, thus avoiding out-of-vocabulary issues and ensuring smooth integration with autoregressive generation tasks.
We formulate a multi-task SFT framework based on LC-Rec, which includes multiple tasks: Sequential Item Prediction, Asymmetric Item Prediction, Item Prediction Based on User Intention, and Personalized Preference Inference. These tasks aim to enhance the model’s ability to capture sequential dependencies, understand implicit user preferences, and align user behavior with items diversely and adaptively.
However, LC-Rec remains limited in capturing user–item collaborative and semantic relationships. To address this, we propose two key enhancements: First, we inject the user's SCID token into all task prompts, as shown in Figure~\ref{fig:fig3}, to ensure richer contextual alignment. Second, we introduce bi-directional alignment objectives (B.2): one task predicts a user's SCID token from their profile text (text $\to$ SCID), while the other reconstructs the user profile text from a given SCID token (SCID $\to$ text).
Compared with prior works, our SFT design enriches user modeling by directly incorporating SCID tokens and explicitly aligns structured and semantic information through bi-directional tasks, providing a stronger foundation for downstream preference optimization.
\subsection{Preference-level Alignment: Progressive DPO with Self-Play and Real-world Feedback}
Although previous stages enable the model to have preliminary recommendation capability, simple preference optimization after SFT is insufficient for continual improvement or robust business alignment, due to the limited coverage of annotated preference data, which fails to capture the full complexity of real recommendation scenarios.
To address this, we introduce progressive DPO with self-play (SP-DPO) and real-world feedback (RF-DPO). Specifically, SP-DPO first leverages self-play to acquire basic generative abilities,  by generating diverse and informative data, thereby mitigating data sparsity and exploration limitations. Then RF-DPO utilizes real-world feedback to constrain the model toward real recommendation tasks, forming a synergistic, progressive learning strategy.
The progressive DPO is based on the Softmax-DPO~\cite{softmax-dpo} by constructing training samples containing multiple rejected responses, which is initialize as SFT model. The training objective for each stage is formally defined as:
\begin{small}
\begin{align}
&\mathcal{L}(\pi_{\theta}^i, \pi_{\mathrm{ref}}^i) 
 = -\mathbb{E}_{\left(x, y_w^i, Y_l^i\right) \sim \mathcal{D}^i} \Bigg[ 
 \log \sigma \Bigg( 
        -\log \sum_{y_l^i \in Y_l^i} \exp \\
        \Bigg( 
          &  \beta \log \frac{\pi_\theta^i\left(y_l^i \mid x\right)}{\pi_{\mathrm{ref}}^i\left(y_l^i \mid x\right)} 
            - \beta \log \frac{\pi_\theta^i\left(y_w^i \mid x\right)}{\pi_{\mathrm{ref}}^i\left(y_w^i \mid x\right)} 
        \Bigg) 
    \Bigg)
    \Bigg]\nonumber 
\end{align}
\label{eqn:3} 
\end{small}
where $\pi_\theta^i$ denotes the current policy at stage $i$, $\pi_{\mathrm{ref}}^i$ is the reference policy, $x$ is the prompt, $y_w^i$ is the chosen response, $Y_l^i$ is the set of rejected responses, $\mathcal{D}^i$ is the progressive training set at stage $i$ and  $\beta$ is a hyperparameter, and $\sigma(\cdot)$ is the sigmoid function. The fine-tuned model at each stage becomes the reference model for the next stage ($\pi_{\theta}^i$ → $\pi_{\mathrm{ref}}^{i+1}$), enabling the model to progressively adapt to more nuanced preference distinctions.
Progressive SP-DPO leverages self-play mechanisms to enhance the model's generative capability by comparing its generated outputs with the ground truth. Specifically, considering the hierarchical nature of SCID, we divide progressive SP-DPO learning into three stages: Easy, Medium, and Hard, using a prefix-ngram match metric~\cite{PNM} (the same prefix exhibits similar semantic and collaborative information). For the easy stage, the chosen and rejected SCID responses of preference data are completely different, with no shared prefix-ngram, making them easy to distinguish. 
For the medium and hard stages, the prefix-ngram overlap between chosen and rejected SCID responses progressively increases, increasing the difficulty of discrimination, but they remain non-identical.
These three-stage preference data, coupled with real user behavior sequences, progressively serve as training data $\mathcal{D}^i$ for preference learning, as shown in Eq~(3). Alternatively, the prefix-ngram match metric can be extended to SCID vector-similarity metric, for a softer sample construction strategy.
Progressive RF-DPO captures authentic user feedback as preference data for alignment by recommending its own generated results to users. Feedback is categorized into three levels: disliked, neutral, and liked. Aligned with progressive learning, training occurs in stages: an easy stage uses strongly disliked items as negatives (liked as positives), while a hard stage uses neutral items as harder negatives (liked remains positive). This staged approach systematically strengthens preference learning. In industrial recommendation settings, levels are defined by user behavior: disliked (explicit negative, e.g., dislike), neutral (implicit negative, e.g., impression without click), liked (positive, e.g., like or purchase). For public datasets (e.g., Amazon reviews), we use an LLM-based sentiment model (ecomgpt~\cite{ecomgpt}) to score reviews, mapping scores to levels: disliked (1), neutral (2-3), and liked (4-5). Integrating this fine-grained feedback enables RF-DPO to better align with user interests and business goals, improving recommendation relevance.
Our progressive DPO framework offers several key advantages. By adopting a progression from easy to hard stages and leveraging both self-play and real-world feedback, the model continually enhances its ability to discern and generalize user preferences, overcoming the “preference ceiling” of static data. The key theoretical insight is that well-designed curricula can provide smoother interpolation between task distributions, leading to more efficient learning compared to direct approaches.
\begin{table*}[t]
\setlength{\abovecaptionskip}{0.05cm}
\setlength{\belowcaptionskip}{0.2cm}
\setlength{\tabcolsep}{3mm}{
\resizebox{\textwidth}{!}{
\begin{tabular}{c|cccc|cccc|cccc}
\toprule
 &
  \multicolumn{4}{c|}{\textbf{Instruments}} &
  \multicolumn{4}{c|}{\textbf{Beauty}} &
  \multicolumn{4}{c}{\textbf{Yelp}} \\
  \multicolumn{1}{c|}{\textbf{Model}} &
  \multicolumn{1}{c}{\textbf{R@5}} &
  \multicolumn{1}{c}{\textbf{R@10}} &
  \multicolumn{1}{c}{\textbf{N@5}} &
  \multicolumn{1}{c|}{\textbf{N@10}} &
  \multicolumn{1}{c}{\textbf{R@5}} &
  \multicolumn{1}{c}{\textbf{R@10}} &
  \multicolumn{1}{c}{\textbf{N@5}} &
  \multicolumn{1}{c|}{\textbf{N@10}} &
  \multicolumn{1}{c}{\textbf{R@5}} &
  \multicolumn{1}{c}{\textbf{R@10}} &
  \multicolumn{1}{c}{\textbf{N@5}} &
  \multicolumn{1}{c}{\textbf{N@10}} \\
\midrule
\multicolumn{13}{c}{\textbf{Traditional Recommendation Methods}} \\ \midrule
\textbf{MF} &
  0.0479 & 0.0735 & 0.0330 & 0.0412 &
  0.0294 & 0.0474 & 0.0145 & 0.0191 &
  0.0220 & 0.0381 & 0.0138 & 0.0190 \\
\textbf{LightGCN} &
  0.0794 & 0.1000 & 0.0662 & 0.0728 &
  0.0305 & 0.0511 & 0.0194 & 0.0260 &
  0.0248 & 0.0407 & 0.0156 & 0.0207 \\
\midrule
\multicolumn{13}{c}{\textbf{Sequential Recommendation Methods}} \\ \midrule
\textbf{Caser} &
  0.0543 & 0.0710 & 0.0355 & 0.0409 &
  0.0205 & 0.0347 & 0.0131 & 0.0176 &
  0.0150 & 0.0263 & 0.0099 & 0.0134 \\
\textbf{HGN} &
  0.0813 & 0.1048 & 0.0668 & 0.0774 &
  0.0325 & 0.0512 & 0.0206 & 0.0266 &
  0.0186 & 0.0326 & 0.0115 & 0.0159 \\
\textbf{Bert4Rec} &
  0.0671 & 0.0822 & 0.0560 & 0.0608 &
  0.0203 & 0.0347 & 0.0124 & 0.0170 &
  0.0186 & 0.0291 & 0.0115 & 0.0159 \\
\textbf{SASRec} &
  0.0751 & 0.0947 & 0.0627 & 0.0690 &
  0.0380 & 0.0588 & 0.0246 & 0.0313 &
  0.0183 & 0.0296 & 0.0116 & 0.0152 \\
\textbf{BigRec} &
  0.0513 & 0.0576 & 0.0470 & 0.0491 &
  0.0243 & 0.0299 & 0.0181 & 0.0198 &
  0.0154 & 0.0169 & 0.0137 & 0.0142 \\
\midrule
\multicolumn{13}{c}{\textbf{Generative and LLM-based Recommendation Methods}} \\ \midrule
\textbf{P5-SemID} &
  0.0775 & 0.0964 & 0.0669 & 0.0730 &
  0.0393 & 0.0584 & 0.0273 & 0.0335 &
  0.0202 & 0.0324 & 0.0131 & 0.0170 \\
\textbf{P5-CID} &
  0.0809 & 0.0987 & 0.0695 & 0.0751 &
  0.0404 & 0.0597 & 0.0284 & 0.0347 &
  0.0219 & 0.0347 & 0.0140 & 0.0181 \\
\textbf{TIGER} &
  0.0870 & 0.1058 & 0.0737 & 0.0797 &
  0.0395 & 0.0610 & 0.0262 & 0.0331 &
  0.0253 & 0.0407 & 0.0164 & 0.0213 \\
\textbf{LETTER-TIGER} &
  {0.0909} & {0.1122} & {0.0763} & {0.0831} &
  {0.0431} & {0.0672} & {0.0286} & {0.0364} &
  {0.0277} & {0.0426} & {0.0184} & {0.0231} \\
\textbf{LC-Rec} &
  0.0824 & 0.1006 & 0.0712 & 0.0772 &
  0.0443 & 0.0642 & 0.0311 & 0.0374 &
  0.0230 & 0.0359 & 0.0158 & 0.0199 \\
\textbf{LETTER-LC-Rec} &
  {0.0913} & {0.1115} & {0.0789} & {0.0854} &
  {0.0505} & {0.0703} & {0.0355} & {0.0418} &
  {0.0255} & {0.0393} & {0.0168} & {0.0211} \\
\textbf{EAGER-LLM} &
  {0.0991} & {0.1224} & {0.0851} & {0.0926} &
  {0.0548} & {0.0830} & {0.0369} & {0.0459} &
  {0.0373} & {0.0569} & {0.0251} & {0.0315} \\
  \textbf{Align$^3$GR} & 
  \textbf{0.1103} & \textbf{0.1442} & \textbf{0.0947} & \textbf{0.1113} & 
  \textbf{0.0627} & \textbf{0.0994} & \textbf{0.0434} & \textbf{0.0529} &
  \textbf{0.0425} & \textbf{0.0679} & \textbf{0.0299} &\textbf{0.0403} \\
  \midrule
\textbf{Improvement} &
  \textbf{+11.3\%} & \textbf{+17.8\%} & \textbf{+11.3\%} & \textbf{+20.2\%} &
  \textbf{+14.4\%} & \textbf{+19.8\%} & \textbf{+17.6\%} & \textbf{+15.3\%} &
  \textbf{+13.9\%} & \textbf{+19.3\%} & \textbf{+19.1\%} & \textbf{+27.9\%} \\
\bottomrule
\end{tabular}
}}
\caption{Overall performance comparison across public datasets.}
\label{tab:MainTable}
\end{table*}
\section{Experiments}
\subsection{Experimental Settings}
\textbf{Datasets.} We conduct experiments on three real-world sequential recommendation datasets from diverse domains: (1) Instruments, a subset of the Amazon review corpus focusing on user interactions with musical equipment; (2) Beauty, also from the Amazon review datasets, contains extensive user behaviors related to beauty products; and (3) Yelp, comprising user-business interactions from the Yelp challenge dataset. For fair comparison, we preprocess the data following standard protocols~\cite{lc-rec, tiger, letter}, filtering users and items with fewer than five interactions and applying the leave-one-out strategy for splitting into training, validation, and test sets. We restrict each user's history length to a maximum of 20 for all sequential models.  Finally, we deploy Align$^3$GR online and conduct A/B tests to further validate its performance on an industrial advertising recommendation platform.
\textbf{Baselines.} We compare our method with strong baselines from both conventional and generative recommender systems, covering various tokenization and alignment strategies:
(1) \textbf{MF}~\cite{mf} (Matrix Factorization); (2) \textbf{Caser}~\cite{caser}; (3) \textbf{HGN}\cite{hgn}; (4) \textbf{BERT4Rec}~\cite{bert4rec}; (5) \textbf{LightGCN}~\cite{lightgcn}; (6) \textbf{SASRec}~\cite{sasrec};
(7) \textbf{BIGRec}~\cite{bigrec}, an LLM-based GR using item titles as textual identifiers;
(8) \textbf{P5-SemID}~\cite{letter}, leveraging item metadata for semantic identifiers;
(9) \textbf{P5-CID}~\cite{letter}, incorporating collaborative signals via clustering for LLM-based models;
(10) \textbf{TIGER}~\cite{tiger}, which applies codebook-based quantized identifiers;
(11) \textbf{LC-Rec}~\cite{lc-rec}, enhancing codebook tokenization with auxiliary alignment tasks;
(12) \textbf{LETTER}~\cite{letter}, a learnable tokenizer for generative recommendation;
(13) \textbf{EAGER-LLM}~\cite{eager-llm}, which further models user-item collaborative signals for token-level alignment.
All baselines are implemented or adapted using open-source code where available.
\textbf{Evaluation Metrics.} We evaluate model performance using standard top-$K$ metrics: Recall@$K$ (R@$K$) and NDCG@$K$ (N@$K$) for $K\in{5, 10}$, following previous work~\cite{tiger, lc-rec,eager-llm}. During training, we limit each user’s historical interaction sequence to the most recent 20 items. For generative methods utilizing beam search, we follow EAGER-LLM and consistently set the beam width to 20.
\textbf{Implementation Details.} Our method is instantiated on Llama2-7B~\cite{llama2} as the backbone LLM, with LoRA-based parameter-efficient fine-tuning~\cite{lora}. For item tokenization, we use a 3-level RQ-VAE, each codebook containing 256 embeddings of dimension 32. SCID representations for both users and items are incorporated into the model vocabulary to prevent OOV.
Training is performed for 20,000 steps using the AdamW optimizer with a batch size of 1024. The learning rate is selected from the set ${1\mathrm{e}{-3},\ 5\mathrm{e}{-4},\ 1\mathrm{e}{-4}}$ based on validation performance. All experiments are conducted on 4 NVIDIA GPUs. Hyperparameters such as $\alpha$ and $\beta$ are tuned on the validation set. For each sample of Softmax-DPO, the number chosen is set to 1, and the number rejected is set to 20. During evaluation, we report the average results over five runs with different random seeds. 
\begin{figure}[h]
    \centering
    \includegraphics[width=1.0\linewidth]{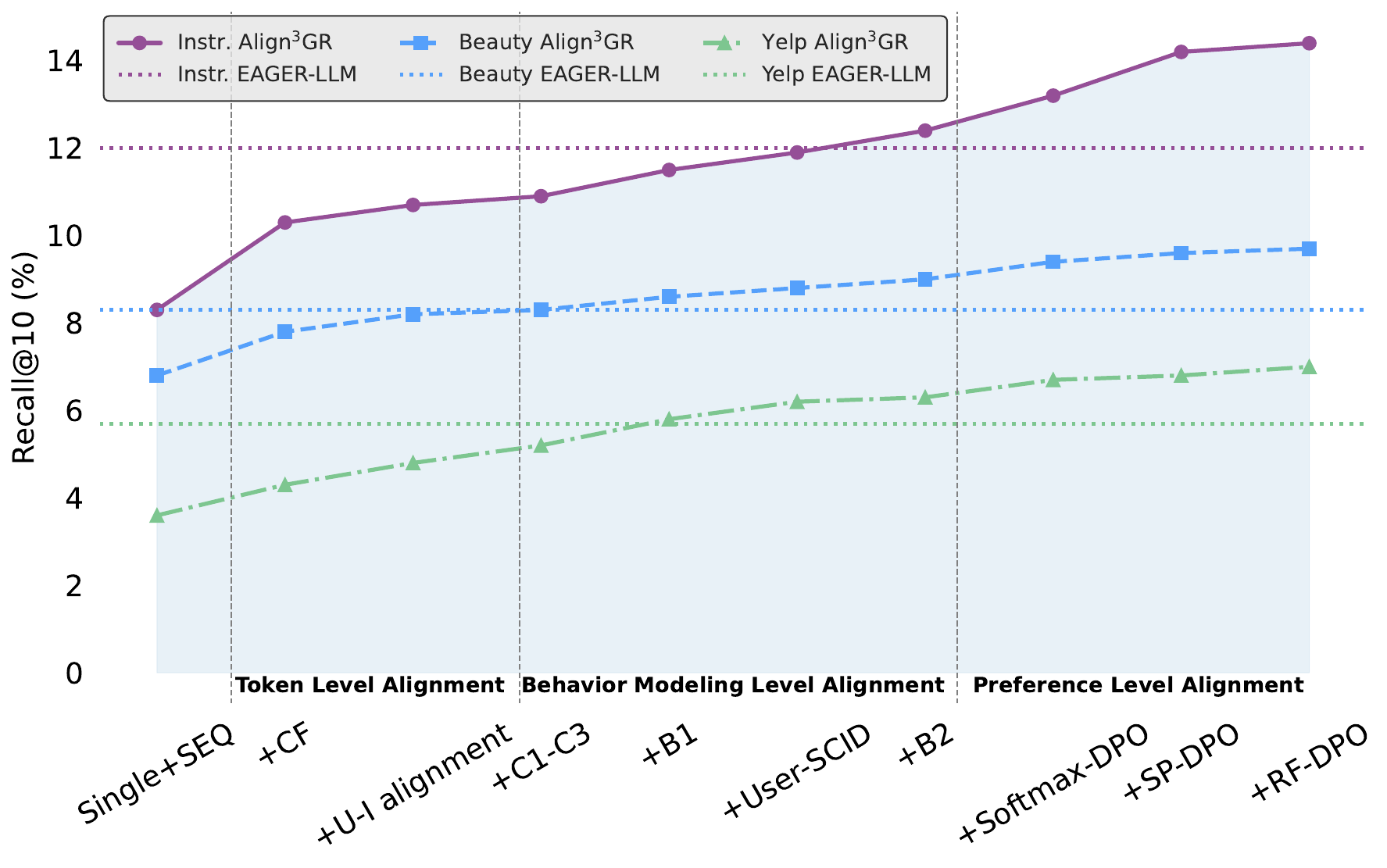}
    \caption{Recall@10 (\%) under incremental alignment configurations; ``Single+SEQ'' denotes using item-side semantic IDs as tokens for the sequence task, while ``+'' indicates cumulative addition of each module.}
    \label{fig:fig4}
\end{figure}
\subsection{Offline Performance}
Table~\ref{tab:MainTable} reports the overall offline performance of all methods on three benchmark datasets. From the results, we have the following observations:
\textbf{First, Align$^3$GR consistently achieves the best or competitive results across all datasets and evaluation metrics}. Compared to the strongest generative baselines such as LETTER and LC-Rec, our method yields substantial improvements with p $<$ 0.05, especially on the Instructments dataset, where it surpasses EAGER-LLM by 17.8\% in Recall@10 and 20.2\% in NDCG@10. This demonstrates the advantage of our multi-level alignment strategy in effectively capturing complex user preferences and collaborative relationships.
\textbf{Second, We visualize the incremental improvements under incremental alignment configurations.} Figure~\ref{fig:fig4} reveals a clear upward trajectory. Notably, replacing item-side semantic IDs with dual learning-based item-side SCIDs yields a sharp performance boost in token-level alignment, underscoring the value of modeling collaborative semantics at the token level.
The preference-level alignment stage yields the most substantial improvement, underscoring the effectiveness of progressive DPO with self-play and real-world feedback.
Throughout all stages, our method significantly outperforms the SOTA baseline (EAGER-LLM), confirming the superiority of multi-level alignment in bridging the gap between LLMs and RS.
\textbf{Finally, these results underscore the effectiveness of Align$^3$GR.} By jointly optimizing user and item representations at the token, behavior, and preference levels, Align$^3$GR is able to capture collaborative signals and dynamic user intents comprehensively. The significant performance gains across offline evaluations confirm that each level of alignment, together with our unified tokenization and adaptive preference optimization strategies, is essential for achieving robust, scalable generative recommendation.
\begin{table}[t]
\centering
\small
\setlength{\tabcolsep}{4.5pt}
\begin{tabular}{l|ccc}
\toprule
~      & \textbf{Baseline} & \textbf{TIGER} & \textbf{Align$^3$GR}    \\
\midrule
\textbf{Recall@100}        & 0.218 & 0.229  & \textbf{0.242}  \\
\textbf{Revenue (Improve.)}        & - & 0.555\%$\textcolor{black}{\uparrow}$   & \textbf{1.432}\% $\textcolor{black}{\uparrow}$  \\
\bottomrule
\end{tabular}
\caption{Performance comparison in industrial scenario.}
\label{tab:industry_data}
\end{table}
\subsection{Online A/B Test}
We further validate the practical efficacy of Align$^3$GR using an internal industrial dataset and online A/B test. On a large-scale industrial advertising recommendation platform, we allocated 10\% of the traffic (approximately 40+ million users) over multiple weeks for A/B testing. As Table~\ref{tab:industry_data} demonstrates, Align$^3$GR surpasses both an industrial two-tower retrieval baseline and the generative TIGER model in online retrieval performance (recall@100) while significantly improving critical business metrics. Specifically, Align$^3$GR achieves a statistically significant \textbf{+1.432\%} revenue improvement in all advertising scenarios under full-scale deployment.
These results confirm that Align$^3$GR's multi-level alignment translates offline advantages into measurable business value in production environments at scale.
\subsection{Ablation Study}
To thoroughly investigate the effect of each alignment level, we conduct comprehensive ablation studies on the Instruments dataset. \textit{To ensure fairness, when ablations are performed on one alignment level, the configurations of the remaining levels are fixed to their best settings.} For example, when studying token-level alignment, we adopt the full setting of SFT tasks and the progressive RF-DPO strategy.
\textbf{Effect of Dual SCID Tokenization.}
Table~\ref{tab:token} presents the ablation results on the Instruments dataset to evaluate the effectiveness of Dual SCID Tokenization.
First, simply switching from single-sided (item-only) to dual-sided tokenization yields a significant performance boost, demonstrating the necessity of jointly modeling both user and item token representations.
Second, incorporating CF on top of semantic inputs further enhances performance under both single and dual settings, highlighting the importance of integrating collaborative signals during representation learning.
Third, enabling U-I alignment via the U2I behavior loss leads to consistent gains, especially when combined with dual tokenization and CF.
This confirms the efficacy of our joint optimization strategy in bridging the gap between user and item SC embeddings.
Overall, these results justify the choices for SCID construction and demonstrate that all three components are complementary and critical for optimal recommendation performance.
\begin{table}[h]
\begin{center}
\small
\setlength{\tabcolsep}{4pt}
\begin{tabular}{ccc|cc}
\toprule
\textbf{Tokenization} & \textbf{CF} & \textbf{U-I Alignment} & \textbf{Recall@10} & \textbf{NDCG@10} \\
\midrule
Item & \xmark   & \xmark & 0.1322 & 0.0978 \\
Item & \cmark   & \xmark & 0.1346 & 0.0991 \\
\midrule
Dual & \xmark   & \xmark & 0.1390 & 0.1032 \\
Dual & \xmark   & \cmark & 0.1426 & 0.1083 \\
Dual & \cmark   & \xmark & 0.1428 & 0.1091 \\
Dual & \cmark   & \cmark & \textbf{0.1442} & \textbf{0.1113} \\
\bottomrule
\end{tabular}
\end{center}
\caption{Ablation study on dual SCID tokenization. Specifically, ``Single'' denotes using only item-side tokenization without user-item joint encoding, while ``Dual'' activates both user and item tokenization. The ``CF'' column indicates whether collaborative features are included, and ``U-I Alignment'' specifies if user and item embeddings are jointly optimized by U2I behavior loss.}
\label{tab:token}
\end{table}
\begin{table}[t]
\centering
\small
\setlength{\tabcolsep}{2.5pt}

\begin{tabular}{l|cccc}
\toprule
\textbf{Methods} &  \textbf{Recall@5} & \textbf{Recall@10} & \textbf{NDCG@5} & \textbf{NDCG@10} \\
\midrule
SEQ               & 0.1042 & 0.1329 & 0.0867 & 0.0982 \\
+ $C_1-C_3$       & 0.1046 & 0.1344 & 0.0881 & 0.0988 \\
+ $B_1$           & 0.1054 & 0.1399 & 0.0908 & 0.1045 \\
\midrule
+ User SCID       & 0.1091 & 0.1417 & 0.0937 & 0.1051 \\
+ $B_2$           & \textbf{0.1103} & \textbf{0.1442} & \textbf{0.0959} & \textbf{0.1113} \\
\bottomrule
\end{tabular}
\caption{Ablation study of various semantic alignment tasks.}
\label{tab:sft}
\end{table}
\textbf{Effect of Behavior Modeling-level Alignment Tasks.}
Table~\ref{tab:sft} presents the ablation results for multi-task SFT on the Instruments dataset. We begin with the SEQ task as the backbone, using the item SCIDs from users' historical behaviors as the sequence.
Introducing user SCID yields additional gains, indicating that structured and informative token representations help the LLM better capture user-item interaction semantics.
Furthermore, the inclusion of user-side alignment ($B_2$), which introduces bidirectional supervision between user profiles and SCID, brings the most significant performance boost across all metrics, highlighting the critical role of exposing LLMs to structured user semantics.
These findings support our hypothesis that effective alignment requires supervision at both the token level and the semantic level, enabling the model to build stronger correspondence between language and recommendation signals.
\begin{table}[t]
\begin{center}
\small
\setlength{\tabcolsep}{2.5pt}
\begin{tabular}{c|cc|cc}
\toprule
\textbf{DPO Variant} & \textbf{Self-Play} & \textbf{Progressive} & \textbf{Recall@10} & \textbf{NDCG@10} \\
\midrule
Softmax-DPO     &  \xmark & \xmark & 0.1295 & 0.0972 \\
\midrule
SP-DPO          &  \cmark & \xmark & 0.1356 & 0.1033 \\
SP-DPO          &  \cmark & \cmark & 0.1396 & 0.1042 \\
\midrule
RF-DPO          &  \cmark & \xmark & 0.1414 & 0.1049 \\
RF-DPO          &  \cmark & \cmark & \textbf{0.1442} & \textbf{0.1113} \\
\bottomrule
\end{tabular}
\end{center}
\caption{Ablation study on preference-level alignment strategies. Using naive Softmax-DPO method as the baseline, chosen: the ground truth next item SCID; rejected: randomly sample 20 generated item SCIDs. The results of RF-DPO are based on progressive SP-DPO, as shown in Figure~\ref{fig:fig2}. The “Progressive” denotes the easy-to-hard learning.
}
\label{tab:ablation_dpo}
\end{table}
\textbf{Effect of Preference-Level Alignment Tasks.}
To assess the impact of our proposed preference-level alignment strategies, we perform an ablation study on SP-DPO and RF-DPO using the Instruments dataset, as shown in Table~\ref{tab:ablation_dpo}. Starting from a well-trained SFT model, we first examine the DPO baseline and then progressively incorporate Self-Play and Progressive strategies within each framework. 
In the SP-DPO branch, adding Self-Play improves Recall@10 from 0.1295 to 0.1356, and further gains are observed when applying progressive learning. In the RF-DPO branch, progressive training consistently outperforms the static variant, culminating in the best overall performance. 
These results demonstrate the complementary benefits of progressive optimization and real feedback integration in aligning LLMs with user preference signals.
\section{Conclusion}
We propose a unified multi-level alignment framework Align$^3$GR that bridges the gap between LLMs and personalized recommendations through dual SCID tokenization, multi-task behavior modeling, and progressive preference optimization. Experiments on real-world datasets show our method outperforms strong baselines. Results highlight the importance of hierarchical alignment for LLM-based GRs, enabling more robust and adaptive personalization. 
\bibliography{aaai}
\end{document}